\documentclass[printer]{tMOP2e}

\begin{document}
\doi{10.1080/0950034YYxxxxxxxx}
\issn{1362-3044}
\issnp{0950-0340}
\jyear{2007}

\title{Suppression of electron spin decoherence in a quantum dot}

\author{Wenxian Zhang$^{\dag}$,
V. V. Dobrovitski$^\dag$$^\ast$\thanks{$^\ast$Corresponding
author. Email: slava@ameslab.gov}, Lea F. Santos$^\ddag$,
Lorenza Viola$^\ddag$, and B. N. Harmon$^\dag$\\
\vspace{6pt} $^\dag$Ames Laboratory, Iowa State University, Ames,
Iowa 50011, United States \\
\vspace{6pt} $^\ddag$Department of Physics and Astronomy,
Dartmouth College, Hanover, New Hampshire 03755, United States}

\markboth{Wenxian Zhang, V. V. Dobrovitski, Lea F. Santos, Lorenza
Viola, and B. N. Harmon}{Suppression of electron spin decoherence in a
quantum dot}

\maketitle

\begin{abstract}
The dominant source of decoherence for an electron spin in a quantum
dot is the hyperfine interaction with the surrounding bath of nuclear
spins.  The decoherence process may be slowed down by subjecting the
electron spin to suitable sequences of external control pulses.  We
investigate the performance of a variety of dynamical decoupling
protocols using exact numerical simulation.  Emphasis is given to
realistic pulse delays and the long-time limit, beyond the domain
where available analytical approaches are guaranteed to work. Our
results show that both deterministic and randomized protocols are
capable to significantly prolong the electron coherence time, even
when using control pulse separations substantially larger than what
expected from the {\em upper cutoff} frequency of the coupling
spectrum between the electron and the nuclear spins.  In a realistic
parameter range, the {\em total width} of such a coupling spectrum
appears to be the physically relevant frequency scale affecting the
overall quality of the decoupling.
\end{abstract}

\section{Introduction}

The dynamics of electron spins in semiconductor structures, such as
quantum dots (QDs), has attracted much attention due to both its
fundamental importance, as well as the potential applications in
spin-based quantum information processing: not only does the electron
spin degrees of freedom provide a natural two-level system with long
energy relaxation time \cite{Elzerman04}, but semiconductor-based
quantum computing architectures can benefit from well-developed
technology and a high degree of scalability \cite{Loss98}. Practical
exploitation of such a potential, however, is severely hindered by the
loss of phase coherence caused by the unavoidable coupling between the
electron spin and its surrounding environment.  For typical QD
operating conditions and geometry, decoherence mechanisms associated
with phonon processes or magnetic spin-orbit interactions are
effectively suppressed, leaving the hyperfine coupling with nuclear
spins as the dominant decoherence channel.  The resulting decoherence
dynamics occurs very rapidly, with the coherence times from ensemble
measurements ($T^\ast_2$) being of the order of tens of nanoseconds
for typical GaAs QDs at experimentally relevant (sub-Kelvin)
temperatures and moderate (sub-Tesla) magnetic fields \cite{dqde05}.

As a result, viable methods for decoherence suppression are of key
relevance.  A number of proposals have been put forward for
increasing the coherence time, for instance by seeking a high
degree of nuclear spin polarization \cite{Atac,Burkard99}, by
narrowing the nuclear spin distribution \cite{Stepa06,domenico},
or by cleverly disentangling the bath from the electron spin via
external control \cite{Lu06}. However, these methods remain
challenging to implement by current or near-term capabilities, and
alternative approaches are still actively sought.

In this context, a promising approach is offered by dynamical
decoupling (DD) methods \cite{ViolaDD}.  Borrowing their inspiration
from coherent averaging techniques in high-resolution NMR spectroscopy
\cite{Mehring,NMRDDBook}, DD schemes aim at removing the effect of
environmental coupling by subjecting the electron spin to suitably
tailored sequences of control pulses. The long correlation time and
distinctively non-Markovian behavior of the nuclear spin reservoir
make the DD approach naturally suited for electron spin coherence
control in QDs, as demonstrated by the recent implementation of
single-pulse Hahn echo experiments \cite{dqde05,Taylor06}, which
succeeded at extending $T_2^\ast$ by two orders of magnitude. Thus, it
is both timely and important to undertake a comprehensive theoretical
study of performance of multi-pulse DD protocols, in order to assess
relative strengths and weaknesses, and identify optimal schemes.  Only
a few protocols have been quantitatively analyzed so far. In addition,
the majority of investigations have focused on the case of QDs
subjected to strong magnetic fields, predicting a coherence
enhancement of at least an order of magnitude in GaAs QDs
\cite{Hahnecho05}.  Still, the experimentally important situation of
low-to-moderate fields is significantly more difficult to analyze due
to the competition between purely adiabatic (energy-conserving)
decoherence and relaxation processes.  Existing studies have mostly
relied on restrictive approximations -- in particular, semi-classical
treatments \cite{Bergli} or Magnus expansion (ME) techniques
\cite{Khodjasteh05}.  Thus, little is known about the actual
performance of DD protocols in a variety of practically important
regimes.

Our goal in this paper is to further push a quantitative analysis of
the electron spin decoherence problem in regimes which are relevant
for experimental DD implementations in QDs, yet have received marginal
consideration so far~\cite{ZhangDD}.  Special attention is devoted to
the limit of zero external field, where dephasing and relaxation must
be simultaneously eliminated, and the asymptotic long-time limit,
where the effects of finite control rate and coherent error
accumulation are crucial, so that neither the ME nor the standard
quasi-static approximation for the nuclear spin bath are reliable a
priori.

The content is organized as follows. Section~\ref{sec:model} provides
the essential background on the QD model Hamiltonian and various DD
protocols.  Section~\ref{sec:res} summarizes both the relevant steps
of the methodology used to tackle the problem, and presents a
comparative analysis of all the DD protocols under investigation.  A
detailed discussion of the best DD protocol based on periodic
concatenated design is presented, including the effects of control
pulse delay, level of concatenation, intra-bath dipolar interaction,
and coupling spectrum.  Concluding remarks are given in
Section~\ref{sec:con}.

\section{Quantum dot model and dynamical decoupling protocols}
\label{sec:model}

The dynamics of an electron spin $S$ coupled to a bath $B$ of $N$
nuclear spins in a QD is described by a total Hamiltonian of the
form \,
\begin{eqnarray}
&H = H_S + H_{SB} + H_B,& \label{eq:h} \\
&H_S = \omega_0 S_z, \hspace{0.5cm} H_{SB} = {\bf S} \cdot
\sum_{k=1}^N A_k {\bf I}_k, \hspace{0.5cm} H_B = \sum_{k>l}^N
\Gamma_{kl} ({\bf I}_k \cdot {\bf I}_l-3I^z_k I^z_l).&
\end{eqnarray}
Here, ${\bf S}$ and ${\bf I}_k$ denote the electron and the $k$-th
bath spin operators, respectively; $\omega_0$ is the electron Zeeman
energy in an external magnetic field $B_0$; $H_{SB}$ gives the
hyperfine contact interaction between the electron spin and the
nuclei; and $H_B$ corresponds to the intra-bath dipolar coupling
between nuclear spins
\cite{Al-Hassanieh06,Dobrovitski03,Zhang06,Taylor06}. The parameters
$A_k$ and $\Gamma_{kl}$ determine the strength of the respective
interactions.  In particular, the coupling energy per nucleus $A_k$
depends on the electron probability density at the nucleus location.
We focus on the limit of zero external field $B_0=0$, so that all
directions are equally affected by decoherence, and choose $I_k=1/2$,
since the analysis does not depend qualitatively on the specific spin
value (up to an appropriate renormalization of the parameters).  The
Zeeman energies of the nuclear spins and the anisotropy of the
hyperfine coupling are very small and can be neglected for the time
scales we are interested in.  The bath is assumed to be initially
unpolarized, which is justified by the fact that in typical
experiments the temperature is much larger than the characteristic
energy of nuclear spins.

In the absence of control pulses, the electron spin described by the
above Hamiltonian undergoes free induction decay (FID) on a time scale
given by $T_2^* = (NA^2/8)^{-1/2}$, where $A=(\sum_k A_k^2/N)^{1/2}$
is approximately $10^{-4}\mu$eV for typical GaAs QDs with $N=10^6$
\cite{Zhang06}. In our analysis, we set $\hbar=1$, and measure
frequency and energy in units of $A$, with time correspondingly being
given in units of $1/A$.

In order to slow down the decoherence process, we recur to DD methods.
The basic idea consists of modifying the system dynamics by subjecting
the electron spin to a sequence of $\pi_{\hat{\bf n}}$ pulses, where
$\hat{\bf n}$ denotes the control axis.  In the simplest setting, the
pulses are equally separated by an interval $\tau$, during which the
system evolves freely. We assume {\em bang-bang controls}, that is,
arbitrarily strong and instantaneous pulses, which we draw from a
control group ${\cal G}=\{ g_j \}$, $j=0,\ldots, |{\cal G}|-1$,
according to $P_{i,j}=g_{j}g_{i}^{\dagger}$. For a single spin as
relevant here, the sequences are derived from the irreducible Pauli
group ${\cal G}_P=\{I,X,Y,Z \}$ \cite{ViolaDD}, where $I, X, Y, Z$ are
the identity and the Pauli matrices of the electron spin,
respectively.  A variety of control protocols have been developed,
which exploit deterministic \cite{ViolaDD,Khodjasteh05} and stochastic
\cite{Viola-random,closed} design.  In {\em cyclic} DD methods, the
pulse sequence is fixed, pre-determined, and periodically repeated in
time; in {\em randomized} DD, the future control actions are not known
in advance, but chosen at random from a set of alternatives.

The construction of deterministic DD sequences has been mainly based
on the average Hamiltonian theory \cite{NMRDDBook}, having as a
goal the cancellation of the dominant terms in the ME series for the
underlying time evolution operator.  The simplest periodic DD (PDD)
sequence we consider here ensures the removal of the zeroth order term
$\overline{\cal H}^{(0)}$ of the average Hamiltonian $\overline {\cal
H}$ at every $T_n=nT_c=n|{\cal G}|\tau$, where $n \in {\mathbb N}$,
$T_c$ is the cycle time, and $|{\cal G}|=4$ is the size of the DD
group.  By fixing the path of ${\cal G}_P$ as $\{I,X,Y,Z \}$, the
control sequence becomes $C_1=C_0XC_0ZC_0XC_0Z$, where $C_0$ denotes
the interval of free evolution.  Thus, the cycle propagator for a
system subjected to the PDD sequence is given by
\begin{eqnarray}
U(T_c) = {\cal T} \exp \left\{-i \int_0^{T_c} [H + H_c(s)] ds \right\}
&\equiv& \exp\left[-i \overline{\cal H} T_c \right] = \exp\left[-i
(\overline{\cal H}^{(0)} + \overline{\cal H}^{(1)} + \cdots)
T_c\right] \label{physical} \\ \nonumber
&=&ZU_0XU_0ZU_0XU_0\,,
\end{eqnarray}
where ${\cal T}$ indicates time ordering, $H_c(t)$ corresponds to
the applied control Hamiltonian, $U_0 = \exp(-iH\tau)$ is the free
propagator, and the zeroth order term in the ME is
\begin{equation}
\overline {\cal H}^{(0)} = \frac{1}{|{\cal G}|}\sum_{k} g_k^{\dagger}
U_0(t_{k+1},t_k) g_k ={1\over 4}\left[ZHZ + YHY + XHX + H\right ] =
H_B\,,
\end{equation}
that is, terms coupling the system and bath degrees of freedom are
removed to lowest order. The terms depending on $H_{SB}$ appear in the
higher-order corrections, in particular $\overline{\cal H} \sim
\overline{\cal H}^{(1)} \propto \tau$, which guarantees DD between the
system and the bath in the limit of arbitrarily fast control, $\tau
\rightarrow 0$. Clearly, $\tau$ is necessarily finite in realistic
settings, its minimum value being constrained by practical limitations
(e.g., for QDs and other devices operating at dilution refrigerator
temperatures, the need to avoid excessive radio-frequency heating of
the sample, which imposes limitation on the total absorbed power).  In
order to improve DD performance, going beyond what is achievable by
the first-order schemes above, two main strategies are available:
higher-order DD, ensuring cancellation of higher order terms in the
effective Hamiltonian; and reduction of error accumulation in time due
to the remaining terms of the Hamiltonian.

As prototypes of higher-level deterministic sequences, we consider
symmetric DD (SDD) and concatenated DD (CDD)~\cite{Khodjasteh05}.  The
SDD sequence corresponds to a symmetrization of $C_1$, being written
as $C_S=C_0XC_0ZC_0XC_0IC_0XC_0ZC_0XC_0$. Its cycle time is twice as
long as PDD, but at every $T_n=2nT_c=8n\tau$, besides $\overline{\cal
H}^{(0)}$, all odd terms in $\overline{\cal H}$ are also canceled,
leading to $\overline{\cal H} \propto \tau^2$.  CDD relies on a
temporal recursive structure, so that at level $\ell+1$ the pulse
sequence is given by
$C_{\ell+1}=C_{\ell}XC_{\ell}ZC_{\ell}XC_{\ell}Z$.  Here, we {\em
truncate} the concatenation procedure at a certain level and repeat a
periodic sequence, referred to as PCDD$_{\ell}$, after every
$4^{\ell}\tau$. For example, $\ell=2$ leads to PCDD$_2$, which is also
a symmetric protocol, although usually superior than SDD at reducing
error accumulation.

The main goal of stochastic protocols is to guarantee that errors
accumulate probabilistically in time, instead of coherently as in
deterministic sequences.  As representatives of these methods, we
consider naive random DD (NRD), random path DD (RPD), and
symmetrized random path DD (SRPD).  NRD corresponds to a sequence
of pulses chosen randomly from ${\cal G}_P$ with equal
probability.  For RPD, randomization is associated with the path
to traverse the group ${\cal G}_P$, that is, at every $T_n=4 \tau$
one chooses at random one of the following options:
\{$C_0XC_0YC_0XC_0Y$, $C_0XC_0ZC_0XC_0Z$, $C_0YC_0XC_0YC_0X$,
$C_0YC_0ZC_0YC_0Z$, $C_0ZC_0XC_0ZC_0X$, $C_0ZC_0YC_0ZC_0Y$\}.
SRPD improves over RPD by symmetrizing the random paths as in SDD
and inserting a random pulse between paths.  Besides slowing down
the error accumulation at long times, RPD and SRPD also ensure
good performance at short times, since they lead to effective
Hamiltonians $H_{eff} \propto \tau $ and $H_{eff} \propto \tau^2
$, respectively.

Note that for randomized DD the periodicity, as well as the notion of
a control cycle, is lost in general.  Thus, the comparison between
deterministic and randomized protocols is most naturally carried out
in the so-called logical frame.  In this frame, $H_c(t)$ is removed
from the evolution through the transformation ${\tilde U}(T) =
U_c^{\dagger} (T) U(T)$, where tilde indicates logical frame, $U_c
(T)={\cal T} \exp [-i \int_0^{T} H_c(s) ds ]$ corresponds to the
control propagator, and $U(T)$ is the propagator in the physical
frame, as given by Eq.~(\ref{physical}).  For deterministic sequences,
both physical and logical propagators stroboscopically coincide at the
completion of each cycle, but the same is not necessarily true for the
stochastic methods.  In the latter case, in order to return the state
of the system to the physical frame, it suffices to keep track of the
applied control pulses.

\section{Performance of control protocols}
\label{sec:res}
\subsection{Control metric}

To compare the above-mentioned control protocols in a way which avoids
dependence on the initial state of the systems, $|\psi_S(0)\rangle$,
we shall consider the worst-case scenario and compute the minimum
input-output pure-state fidelity,
\begin{eqnarray}
F_m(T)=\mbox{min}_{|\psi_S(0)\rangle} {\rm Tr} [\tilde{\rho}_S (T) \rho_S(0)],
\end{eqnarray}
where $\tilde{\rho}_S (T)$ corresponds to the reduced density matrix
of the electron spin at time $T$.  Equivalently, $F_m(T)$ may be
thought of as the minimum state overlap between the intended and the
decoupled system evolution.

Analytical bounds on the worst-case pure-state expected fidelity have
been established for various DD
protocols~\cite{Viola-random,closed,Khodjasteh05}, however they apply
only to short evolution times and small pulse separations.  Numerical
simulations are mandatory to investigate DD performance at long times
and for a broad range of $\tau$ values.  We are especially interested
in pulse delays beyond the restricted domain where the ME is {\em
guaranteed} to converge, $\omega_c T_c \ll 1$, where $\omega_c$
denotes a high frequency cutoff.  While in most situations $\omega_c$
is determined by $H_B$, for the QD problem the nuclear spin bath is to
a very high accuracy {\em non-dynamical}, thus $\omega_c$ is
determined by the electron-nuclear coupling Hamiltonian $H_{SB}$.
This leads to $\omega_c\approx \sum_k |A_k|/4 \sim N A/4$, which is
very large for typical QDs.  We will consider $\tau \sim 1/2\sigma$,
where $\sigma$ represents the power spectrum width of the entire
system, with $2\sigma \approx (\sum_k A_k^2)^{1/2}= \sqrt{N} A$
\cite{Melikidze04}.  Thus, the basic cycle time $T_c= 4\tau \sim
\sqrt{N}\omega_c^{-1}$ will be {\em up to a factor $\sqrt N$ longer}
than the formal (worst-case) convergence requirement of the ME.  We
shall also assume that $A_k >0$ are uniformly distributed random
numbers.

The minimum fidelity is numerically obtained by employing the quantum
process tomography method~\cite{Nielsen02} as follows.  The initial
state of the entire system is taken as a direct product of the
electron spin state $|\psi _S(0)\rangle$ and the bath spin state
$|\psi_B(0)\rangle$.  Four different $|\psi _S(0)\rangle$ are
considered: $|\uparrow \rangle, |\downarrow \rangle, (|\uparrow
\rangle + |\downarrow \rangle)/\sqrt{2}$, and $(|\uparrow \rangle +
i|\downarrow \rangle)/\sqrt{2}$; while a single $|\psi_B(0)\rangle$ is
assumed, corresponding to a random superposition $|\psi_B(0)\rangle =
\sum_{i=1}^{2^N} c_i |\phi_i \rangle$ of all possible tensor products
of the form $|\phi \rangle=|\uparrow\rangle_1 \otimes
|\downarrow\rangle_2 \otimes |\downarrow\rangle_3 \otimes \ldots
\otimes |\uparrow\rangle_N$, where $c_i$ are uniform random
numbers. The time-dependent Schr\"odinger equation for the joint
system plus bath dynamics is then solved by applying the Chebyshev
polynomial expansion method to the evolution operator
\cite{Dobrovitski03}.  At final time $T$, a partial trace over the
bath is performed, and the four resulting reduced density matrices are
used to compute the $4\times 4$ $\chi$-matrix~\cite{Nielsen02}. The
$\chi$-matrix is a superoperator that describes the effective dynamics
of the electron spin and allows for the reduced density matrix of an
arbitrary electron initial state to be evaluated from the equation
\begin{eqnarray}
\tilde{\rho}_S(T) &=& \sum_{m,n=0}^3
K_m \rho_S(0) K_n^\dag \, \chi_{mn}(T)\,,
\end{eqnarray}
where $K_0=I$, $K_1=X$, $ K_2=-iY$, and $K_3=Z$.  Since an arbitrary
initial state of $S$ in the Bloch sphere may be expressed as
$|\psi_S(0)\rangle = \cos(\theta/2) |\uparrow \rangle
+\sin(\theta/2)e^{i\varphi}|\downarrow \rangle $, $\theta \in
[0,\pi]$, $\varphi \in [0,2\pi]$, $F_m(T)$ simplifies to
\begin{eqnarray}
F_m(T)=\mbox{min}_{(\theta, \varphi)} \,\langle \psi_S(0)|\tilde{\rho}_S
(T)|\psi_S(0)\rangle \,.
\end{eqnarray}
In practice, after determining the matrix $\chi$, we obtain $F_m(T)$
from numerical minimization, by comparing the results over a
statistically meaningful set of initial guesses for
$(\theta,\varphi)$, and then selecting the worst case.

\subsection{Results}

\begin{figure}[h]
\begin{center}
\includegraphics[width=4in]{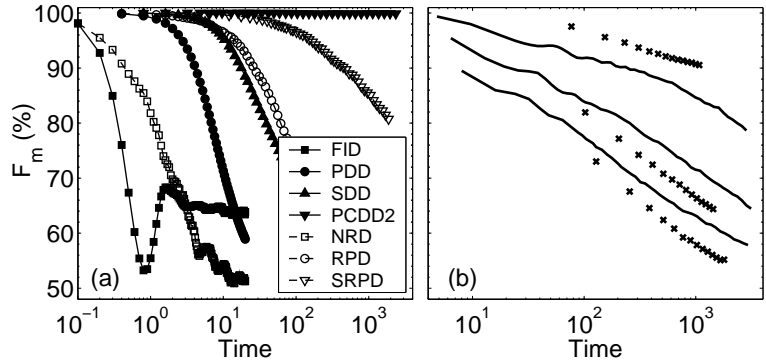}
\end{center}
\caption{Minimum fidelity vs. time in the logical frame. The
Hamiltonian parameters are $B_0=0$, $\Gamma_0=0$, and $N=15$.  (a)
Comparison of various protocols with $\tau=0.1$.  For NRD and FID,
data points are acquired after every $\tau$, for deterministic DD
(PDD, SDD, PCDD$_2$), right at the completion of each cycle, while for
RPD and SRPD this is done after every $4\tau$ and $8\tau$,
respectively.  Random protocols are averaged over $10^2$ control
realizations.  (b) Comparison of PCDD$_2$ (solid lines) and PCDD$_4$
(crosses) for $\tau=0.3$, $0.4$, $0.5$ from top to bottom.  }
\label{fig:cp}
\end{figure}

Fig.~\ref{fig:cp}(a) shows $F_m(T)$ for the above-mentioned DD
protocols. For comparison, we plot also the FID signal, for which no
control is applied.  Because the characteristic time scale $\tau_D$
for the nuclear dipolar dynamics (resulting from $H_B$) is at least
two orders of magnitude slower than the one due to $H_{SB}$ in typical
QDs, setting $H_B=0$ is justified over the time scales studied here.
As seen, all schemes lead to substantial enhancement of the electron
spin coherence, provided that the control time scale is sufficiently
fast compared to an {\em effective upper cutoff} determined by the
spectral width $2\sigma$.  Similar to the behavior observed for DD of
$1/f$ noise \cite{fnoise}, this indicates how general convergence
arguments may place too stringent a bound for realistic systems,
actual performance depending sensitively on spectral details.

Among the various protocols, NRD shows (not surprisingly) the worst
performance, consistent with the expectation that its advantages over
deterministic DD appear only when the control group is very large.
PDD and SDD are, respectively, outperformed by RPD and SRPD,
confirming the reduction of error accumulation associated with
randomization.  Interestingly, the RPD protocol provides worse
fidelity than SDD at times smaller than $10$, but eventually also
outperforms SDD at longer times.  The best performance for this
system, however, is achieved by the deterministic PCDD$_2$.  Compared
with the case of closed systems under the action of reducible control
groups investigated in~\cite{closed}, this confirms how concatenated
design is especially efficient for irreducible averaging.  A feature
which also proves beneficial for concatenation is the absence of
internal bath dynamics~\cite{Khodjasteh05}.

The above findings prompt a more in-depth analysis of the PCDD
protocol.  In Fig.~\ref{fig:cp}(b), we compare the performance of
two levels of concatenation, $\ell=2,4$, for different values of
$\tau$. As expected, the quality deteriorates as $\tau$ increases
but, unexpectedly, PCDD$_4$ underperforms PCDD$_2$ for larger
$\tau$, suggesting that predictions based on applications of the
ME~\cite{Khodjasteh05} may not hold in this case.

\begin{figure}[t]
\begin{center}
\includegraphics[width=5in]{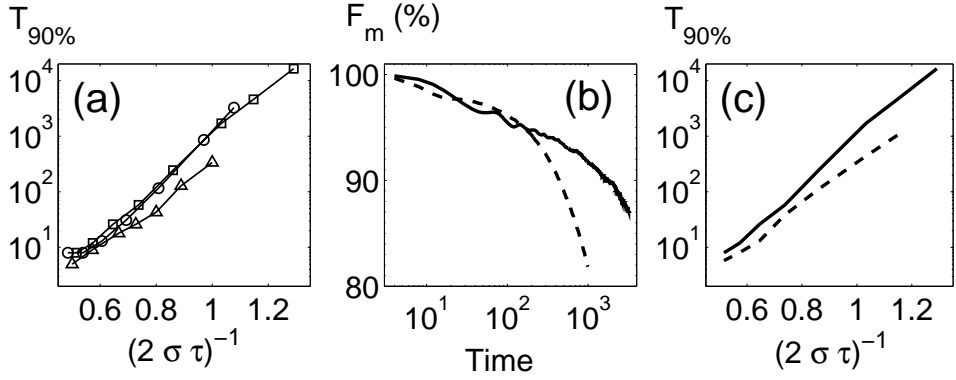}
\end{center}
\caption{(a) $T_{90\%}$ vs. $\tau $: Comparison of PCDD$_2$ for bath
sizes $N=15$ (squares), $17$ (circles), and $25$ (triangles);
$H_B=0$. The time $T_{90\%}$ depends strongly on $\tau$, varying by
three orders of magnitude when $\tau$ is varied by a factor of
two. (b) The effect of dipolar interaction between nuclear spins on
PCDD$_2$ performance, $F_m$ {\it vs.} time for $\tau=0.25$; (c)
$T_{90\%}$ {\it vs.} $\tau$. For both (b) and (c), the parameters are
$N=15$, $B_0=0$, and uniformly distributed random numbers $\Gamma_{kl}
\in [-\Gamma_0,\Gamma_0]$; $\Gamma_0=0$ (solid line), $\Gamma_0=0.1$
(dashed line).} \label{fig:dip}
\end{figure}

The dependence of the PCDD$_2$ performance on the pulse delay
$\tau$ is investigated in Fig.~2(a), by focusing on the time
$T_{90\%}$ where the protocol attains $F_m(T_{90\%})=0.9$ for
given $\tau$.  Two main features are observed: (i) $T_{90\%}$
increases monotonically as $\tau$ decreases -- consistent with the
fact that no crossover from decoherence suppression to decoherence
acceleration is expected as long as $\tau$ is sufficiently small
with respect to the inverse spectral width of the environmental
coupling \cite{ViolaDD,fnoise}. That is, in terms of the formal
analogy established between decoherence control methods based on
DD pulses and frequent quantum
measurements~\cite{Kofman,Facchi04}, only the counterpart of
quantum Zeno effect (QZE) is observed in the relevant parameter
regime, with no occurrence of ``anti-Zeno'' behavior; (ii) The
PCCD$_2$ performance depends quite sensitively on $\tau$, moderate
changes in the value of $\tau$ (say, by a factor of two) changing
the coherence time $T_{90\%}$ by up to three orders of magnitude.
For PCDD$_2$, the leading non-zero term in the ME is
$\overline{\cal H}^{(2)}$, so that the error amplitude per cycle
is bounded by ${\cal O}( |{\cal G}|^3 (\sigma \tau)^3)$, and
therefore the total error amplitude after a number
$T_{90\%}/(|{\cal G}| \tau)$ of cycles grows as ${\cal O}(T_{90\%}
\sigma^3 ( |{\cal G}|\tau)^2 )$.  As a result, within the regime
of validity of the ME, the time $T_{90\%}$ at which the total
error probability $1-F_m(T_{90\%})$ reaches $10\%$ is expected to
increase as $1/\tau^2$ with decreasing $\tau$.  The steeper
dependence shown by the data in Fig.~2(a) indicates that a naive
extrapolation of the above estimate in the long-time regime of
relevance is not directly viable.

Finally, we investigate the effect of interaction between the bath
spins, taking the intra-bath coupling parameters $\Gamma_{kl}$ in
(\ref{eq:h}) as uniformly distributed random numbers in
$[-\Gamma_0, \Gamma_0]$, and assuming that the bath spins are
located on a $3 \times 5$ piece of a two-dimension square lattice.
One may expect that the effect of intra-bath coupling becomes
important once the characteristic coupling strength $\Gamma_0$ is
comparable with the system-bath coupling $A$. However,
Figs.~\ref{fig:dip}(b) and (c) show that the effect of intra-bath
interactions becomes clearly visible already for much smaller
values of $\Gamma_0=0.1 A$, leading to a decrease of the coherence
time $T_{90\%}$ by an order of magnitude in comparison with the
case $\Gamma_0=0$. Although the fast bath dynamics regime is not
directly relevant to standard GaAs QDs, where $\Gamma_0\sim 0.01
A$, our observations may be useful as a first step towards
characterization of DD in other systems.

\subsection{Coupling spectrum}

\begin{figure}[t]
\begin{center}
\includegraphics[width=4.15in]{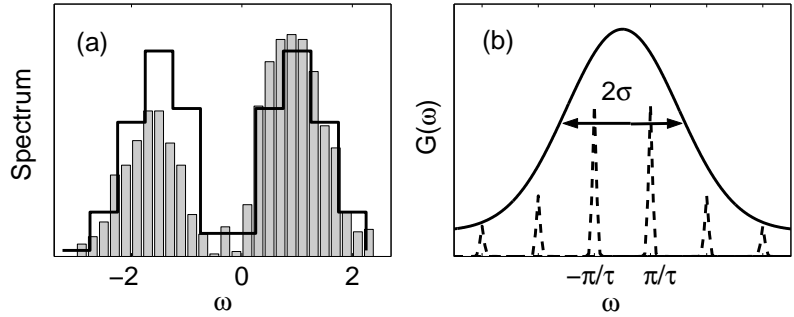}
\end{center}
\caption{(a) Spectrum of the QD for $B_0=0$, $\Gamma_0=0$, and
$N=10$. The bars are obtained for a single realization of random
$A_k$, whereas the solid line corresponds to the homogeneous limit,
$A_k=A$. (b) Pictorial representation of the coupling spectrum
$G(\omega)$ (solid line) and the control pulse spectrum (dashed
lines).}
\label{fig:spectrum}
\end{figure}

Additional physical insight into the performance of DD schemes may be
gained by considering the coupling spectrum of the QD Hamiltonian
(\ref{eq:h}) in the absence and in the presence of control. A typical
spectrum for uniformly distributed random $A_k$ (obtained numerically)
is shown in Fig.~\ref{fig:spectrum}(a). For the homogeneous case of
constant $A_k=A$ (below, we set $A=1$ for convenience), the spectrum
may be evaluated analytically, the result being also shown in
Fig.~\ref{fig:spectrum}(a). Physically, the limit $A_k=A$ is
equivalent to a semiclassical treatment of the bath, where the
quantum-mechanical operator ${\bf B}=\sum_k A_k {\bf I_k}$,
representing the Overhauser field exerted by the bath on the electron
spin, is approximated by a randomly oriented classical vector ${\bf
B}$ with zero mean and variance $\sigma$
\cite{Bergli,Zhang06,Taylor06}.

For $A_k=A$, the operator of the total bath spin ${\bf I} = \sum_k
{\bf I}_k$ is an integral of motion, and the Hamiltonian (\ref{eq:h})
acquires a simple form of two coupled spins,
\begin{equation}
H=A{\bf S}\cdot {\bf I}=H_{diag}+ H_{off}= A\, S_z I_z + A(S_xI_x +
S_yI_y) \,,
\label{h2s}
\end{equation}
where $I\gg 1$.  Let us consider a state of the whole system
$|\uparrow;I, I_z\rangle$, which corresponds to the electron spin
up, and the bath having total spin $I$ with $z$-projection equal
to $I_z$. Since the value of $I_z+S_z$ is also an integral of
motion, this state is coupled only to one state, $|\downarrow;I,
I_z+1\rangle$.  Thus, the total Hamiltonian matrix can be
separated into a number of uncoupled $2\times 2$ blocks, each
block involving a pair of levels $|\uparrow;I, I_z\rangle$ and
$|\downarrow;I, I_z+1\rangle$. Correspondingly, the whole system
can be viewed as a collection of uncoupled two-level systems, with
state space spanned by basis states $|\uparrow;I, I_z\rangle$ and
$|\downarrow;I, I_z+1\rangle$. The energy splitting between these
basis states, $\epsilon=I_z+1/2$, is determined by the diagonal
part of the Hamiltonian (\ref{h2s}), $H_{diag}$, whereas the
coupling matrix element between the basis states is determined by
the non-diagonal part $H_{off}$.

The coupling spectrum $G(\omega)$ quantifies the density of bath
states weighted by the corresponding coupling strength.  The spectral
density $n(\omega)$ of two-level systems having energy splitting
$\epsilon=\omega$ may be computed by a combinatorial argument: this is
the density of the bath states having a given value of $I_z$ (since
$\epsilon=I_z$ for $I_z\gg 1$), which leads to
\begin{equation}
n(\omega) = {1\over \sqrt{2\pi}\;\sigma} \exp(-\omega^2/2\sigma^2).
\end{equation}
The modulus square of the matrix element $V$ which couples the basis
states $|\uparrow;I, I_z\rangle$ and $|\downarrow;I, I_z+1\rangle$ is
$|V|^2=(A^2/4)(I^2-I_z^2)$, and it should be averaged over all
two-level systems having the same $I_z$ (i.e., the same $\omega$) but
different values of $I$.  We then find that $|V|^2 = \sigma^2/2$, and
the coupling spectrum $G(\omega)$ for the electron spin plus bath is
(to a good accuracy) a sum over all coupling spectra of the individual
two-level systems~\cite{Remark},
\begin{equation}
G(\omega) = |V|^2 n(\omega) = {\sigma\over
2\sqrt{2\pi}}\exp(-\omega^2/2\sigma^2). \label{g}
\end{equation}

Control modifies, in general, the above coupling spectrum.  Under DD,
in particular, the relevant spectrum becomes a convolution of the
original coupling spectrum $G(\omega)$ with the spectrum of the
control pulses (Fig.~\ref{fig:spectrum}(b)), which is given by a
series of $\delta$-like peaks separated by a distance of the order of
$1/\tau$ \cite{Kofman}.  The two central peaks of the controlled
spectrum have the weight of $4/\pi^2$ \cite{Kofman}, while the
contribution from the remaining peaks is less than 20\% and thus may
be omitted.  As a result, in the presence of DD pulses, the decay is
caused predominantly by the part of $G(\omega)$ which has $\omega\sim
\pi/\tau$.  Such a contribution decreases exponentially, as
$\exp(-\omega^2/2\sigma^2)$, with decreasing $\tau$.  Correspondingly,
DD becomes efficient for $\tau\sim 1/\sigma$, the inverse width of the
spectrum, not the inverse spectrum cutoff frequency $\omega_c$ (which
is larger than $\sigma$ by a factor of $\sqrt{N} \gg 1$).

The fast decrease of $G(\omega)$ with $\omega$ may provide an
intuition as to why $T_{90\%}$ depends so drastically on $\tau$, as
seen in Fig.~2(a).  If we assume, in analogy with
\cite{fnoise,Kofman}, that the time scale $T_{90\%}$ is inversely
proportional to the contribution of $G(\omega)$ at $\omega=q\pi/\tau$
(where $q$ is some protocol-dependent parameter of the order of
unity), then a strong dependence on $\tau$ is indeed expected:
\begin{equation}
T_{90\%}\sim
{2\sqrt{2\pi}\over\sigma}\exp{(q^2\pi^2/2\sigma^2\tau^2)}.
\end{equation}
Also, Eq.~(\ref{g}) further demonstrates why the anti-Zeno-like
behavior (acceleration of decoherence by DD pulses) is not observed in
our simulations: $G(\omega)$ monotonically decreases with $\omega$,
and even for large $T_c$ the total power of the modified coupling
spectrum cannot increase.  Since, as noted, our results for $N=15$,
$17$, and $25$ (Fig.~2(a)) are very close to each other, we expect
that the basic qualitative features of our analysis remain applicable
to realistic situations with large $N$.

\section{Conclusion}
\label{sec:con}

We have characterized quantitatively the performance of various DD
protocols for an electron spin coupled to a nuclear spin bath,
including both standard cyclic and randomized DD design.  Two main
points emerge from our analysis: on one hand, DD methods demonstrate a
clear potential for significantly enhancing the electron spin
coherence time for an {\em arbitrary initial state}, actual
performance depending on both control and physical parameters.  In
particular, the frequency spectrum of the environmental coupling
proves fundamental for identifying the {\em minimum} time scale
requirements that must be met in order for DD to be effective.  On the
other hand, care must be taken in extrapolating simplified approaches
(such as the ME) to parameter regimes outside the established domain
of applicability, in particular for coherence preservation over long
times.  While additional investigation is needed to gain a complete
picture (including, for instance, the effect of realistic control
imperfections), it is our hope that the steady progress witnessed in
experimental solid-state control techniques \cite{Chen04,Fraval05} may
soon allow to validate the usefulness of DD in realistic QD devices.

\section{Acknowledgement}

It is a pleasure to thank D. G. Cory, A. Imamoglu, J. J. Longdell,
and A. J. Rimberg for discussions. This work was supported by the
Department of Energy, Basic Energy Sciences, under Contract No.
DE-AC02-07CH11358.  Work at Ames was also supported by the NSA and
ARDA under ARO contract DAAD 19-03-1-0132. L. V. also acknowledges
partial support from the NSF through Grant No. PHY-0555417.


\begin{thebibliography}{12}
\bibitem{Remark} This expression may not apply for extremely small
$\omega$, but the contribution from such part of the spectrum is
negligible.
\end{thebibliography}
\end{document}